\documentclass[preprint,aps,showpacs]{revtex4}
\usepackage{mathrsfs}
\usepackage{amsmath}
\usepackage{amssymb}
\usepackage{epsfig}
\usepackage{graphicx}
\usepackage{booktabs}
\usepackage{array}
\usepackage{multirow}
\begin{document}
\renewcommand{\arraystretch}{0.5}
\newcommand{\beq}{\begin{eqnarray}}
\newcommand{\eeq}{\end{eqnarray}}
\newcommand{\non}{\nonumber\\ }

\newcommand{\acp}{ {\cal A}_{CP} }
\newcommand{\psl}{ p \hspace{-1.8truemm}/ }
\newcommand{\nsl}{ n \hspace{-2.2truemm}/ }
\newcommand{\vsl}{ v \hspace{-2.2truemm}/ }
\newcommand{\epsl}{\epsilon \hspace{-1.8truemm}/\,  }

\def \cpl{ Chin. Phys. Lett.  }
\def \ctp{ Commun. Theor. Phys.  }
\def \epjc{ Eur. Phys. J. C }
\def \jpg{  J. Phys. G }
\def \npb{  Nucl. Phys. B }
\def \plb{  Phys. Lett. B }
\def \prd{  Phys. Rev. D }
\def \prl{  Phys. Rev. Lett.  }
\def \zpc{  Z. Phys. C }
\def \jhep{ J. High Energy Phys.  }

\title{The study of the nonleptonic two body $B$ decays involving a light tensor meson in final states}
\author{Zhi-Tian Zou$^a$\footnote{zouzt@ytu.edu.cn } and  Cai-Dian L\"u $^b$\footnote{lucd@ihep.ac.cn} }
\affiliation{ a. Department of Physics, Yantai University, Yantai 264005, People's Republic of China\\
b. Institute  of  High  Energy  Physics  and  Theoretical  Physics Center for Science Facilities,
Chinese Academy of Sciences, Beijing 100049, People's Republic of China
}

\date{\today}
\begin{abstract}
The nonleptonic two body $B_{u,d,s,c}$ decays involving a light tensor meson in final states are studied
 in the perturbative QCD approach based on $k_T$ factorization. The decay modes
 with a tensor meson emitted,  are prohibited in naive factorization,  since the emission diagram with a tensor meson produced from vacuum is vanished. While  contributions from the so-called
 hard scattering emission diagrams and annihilation type diagrams are important and calculable in the perturbative QCD approach.
 The branching rations of most decays are in the range of $10^{-4}$ to $10^{-8}$, which are bigger by 1 or 2 orders
 of magnitude than the predictions given by naive factorization, but consistent with the predictions from QCD factorization
 and the recent experimental measurements. We also give
 the predictions for the direct $CP$ asymmetries, some of which are large enough for the future experiments to measure.
We also find that,  even with a small mixing angle,
 the mixing   between $f_2$ and $f_2^{\prime}$ can bring remarkable changes to both branching ratios and the direct $CP$ asymmetries for some decays involving $f_2^{(\prime)}$ mesons.
 For   decays with a vector meson and a tensor meson in final states, we predict a large percentage of transverse polarization contributions
 due to the contributions of the orbital angular momentum of the tensor mesons.
\end{abstract}



\maketitle

\section{Introduction}

The two body hadronic decays of $B$ meson are important,  since they can provide constraints of the standard
model Cabibbo-Kobayashi-Maskawa (CKM) matrix, a test of the QCD factorization, information on the decay mechanism and
  the final state interaction, and   also a good place to study $CP$ violation  and     new physics signal.
Most of the studies concentrate on the $B\to PP$, $PV$ and $VV$ decays, since they are easy to be measured at experiments.
Recently, more and more experimental measurements about $B$ decays with a light tensor meson involved in the final states have been obtained\cite{pdg}. Inspired by these experiments, many theoretical studies on the  $B_{u,d,s,c}$ decays involving
a light tensor meson have been done based on naive factorization \cite{prd491645,prd555581,prd59077504,epjc22683,epjc22695,prd67014002,
jpg36095004,arxiv1010.3077,prd67014011,isgw2,isgw3}, QCD factorization (QCDF) \cite{zheng1,zheng2} and perturbative QCD (PQCD) approach
\cite{wwprd83014008,zoupt,zoudt,zoubcdt}.

 The tensor meson emission diagrams are prohibited,
because the amplitude proportional to the matrix element $\langle T\mid j^{\mu}\mid 0\rangle$ vanishes from Lorentz
covariance considerations, where $j^{\mu}$ donates the $(V\pm A)$ current or $(S\pm P)$ density \cite{zheng1,zheng2,epjc22683,epjc22695}.
Thus, these decays modes with a tensor meson emitted are prohibited in naive factorization. On the other hand,
the naive factorization can not give creditable predictions for those color-suppressed or penguin dominant decays. What is more, the naive factorization can not deal with the pure annihilation type decays.
The recent developed QCDF approach
\cite{zheng1,zheng2} and the  PQCD factorization approach
 \cite{zoupt,zoudt,zoubcdt} overcome these shortcomings by including the large
hard scattering contributions and the annihilation type
contributions to give more reliable predictions for these decays, especially those with tensor meson emitted or pure annihilation type decays.
It is worth of mentioning that the annihilation type diagrams can be perturbatively calculated
without parametrization in PQCD approach \cite{annihilation1,annihilation2}.
Only the PQCD approach have successfully predicted the pure annihilation type decays $B_{s}\rightarrow \pi^{+}\pi^{-}$ \cite{prd70034009,prd76074018}
 and $B^0\rightarrow D_{s}^{-}K^{+}$ \cite{annihilation1,prd78014018}, which have been confirmed by experiments later \cite{10498,pdg}.
 So, for those annihilation diagram dominant or pure annihilation type decays, for example, $B_c\rightarrow D^{(*)}T$ decays,
   the calculation in PQCD approach is more reliable than other approaches.

The studied tensor mesons include the isovector $a_2(1320)$, the isodoublet
 $K_{2}^{*}(1430)$, and isosinglet $f_{2}(1270)$ and
 $f_{2}^{\prime}(1525)$ \cite{pdg}. For the tensor meson with $J^{p}=2^{+}$, both the orbital
 angular momentum $L$ and the total spin $S$ of the quark pair are equal
 to $1$. However, their production property in $B$ decays is
 quite similar to the light vector mesons \cite{wwprd83014008}.
  Because of the Bose statistics, the light-cone distribution amplitudes  of tensor mesons are
antisymmetric under the interchange of momentum fractions of the
quark and anti-quark in the flavor SU(3)  limit
\cite{zheng1,zheng2}.

$B$ meson decays into tensor mesons are of prime interest in several aspects. The branching ratios and $CP$ asymmetries
are helpful to inspect those different theoretical calculations. Some decay modes, like $B \to K_2^*(1430)\omega$, possess
a large isospin violation\cite{babar}. Moreover, from our computations, polarizations of the final state mesons in $B$ decays into
 tensor mesons and vector mesons  are also beyond the naive hierarchy, which is similar to some decays to two light vector mesons
  and shed light on the helicity structure of the electroweak interactions.

\section{FORMALISM }\label{sec:function}

It is well known that the key step to predict two-body hadronic $B_{(s,c)}$ decays is calculating the transition matrix elements:
\begin{eqnarray}
\mathcal{M}\propto \langle h_1h_2|\mathcal{H}_{eff}|B_{(s,c)}\rangle
\end{eqnarray}
with the weak effective Hamiltonian $\mathcal{H}_{eff}$ written as \cite{rmp681125}
\begin{eqnarray}
\mathcal{H}_{eff}=&&\frac{G_{F}}{\sqrt{2}}\left\{ V_{u(c)b}^{*}V_{u(c)X}\left[C_{1}(\mu)O_{1}^{q}(\mu)+C_{2}(\mu)O_{2}^{q}(\mu)\right]\right.\nonumber\\
&&\left. -V_{tb}^{*}V_{tX}\left[\sum_{i=3}^{10}C_{i}(\mu)O_{i}(\mu)\right]\right\},
\label{eq:heff}
\end{eqnarray}
with the CKM matrix elements $V_{u(c)b(X)}$ and $V_{tb(X)}$  ($X=d,s$).  $c_{i}(\mu)$are the effective Wilson coefficients at the
 renormalization scale $\mu$, whose expression can be found at ref.\cite{pqcd1}. The local four-quark operators $O_{j}\,(j=1,...,10)$ can be given as
\begin{eqnarray}
O_{1}^{u}=(\bar{b}_{\alpha}u(c)_{\beta})_{V-A}(\bar{u}(\bar{c})_{\beta}X_{\alpha})_{V-A},\;\;\;O_{2}^{q}=(\bar{b}_{\alpha}u(c)_{\alpha})
_{V-A}(\bar{u}(\bar{c})_{\beta}X_{\beta})_{V-A},
\end{eqnarray}
for the current-current (tree) operators,
\begin{eqnarray}
&&O_{3}=(\bar{b}_{\alpha}X_{\alpha})_{V-A}\sum_{q^{\prime}}(\bar{q}^{\prime}_{\beta}q^{\prime}_{\beta})_{V-A},\;\;\;
O_{4}=(\bar{b}_{\alpha}X_{\beta})_{V-A}\sum_{q^{\prime}}(\bar{q}^{\prime}_{\beta}q^{\prime}_{\alpha})_{V-A},\\
&&O_{5}=(\bar{b}_{\alpha}X_{\alpha})_{V-A}\sum_{q^{\prime}}(\bar{q}^{\prime}_{\beta}q^{\prime}_{\beta})_{V+A},\;\;\;
O_{6}=(\bar{b}_{\alpha}X_{\beta})_{V-A}\sum_{q^{\prime}}(\bar{q}^{\prime}_{\beta}q^{\prime}_{\alpha})_{V+A},
\end{eqnarray}
for the QCD penguin operators, and
\begin{eqnarray}
&&O_{7}=\frac{3}{2}(\bar{b}_{\alpha}X_{\alpha})_{V-A}\sum_{q^{\prime}}e_{q^{\prime}}(\bar{q}^{\prime}_{\beta}q^{\prime}_{\beta})_{V+A},
\;\;O_{8}=\frac{3}{2}(\bar{b}_{\alpha}X_{\beta})_{V-A}\sum_{q^{\prime}}e_{q^{\prime}}(\bar{q}^{\prime}_{\beta}q^{\prime}_{\alpha})_{V+A},\\
&&O_{9}=\frac{3}{2}(\bar{b}_{\alpha}X_{\alpha})_{V-A}\sum_{q^{\prime}}e_{q^{\prime}}(\bar{q}^{\prime}_{\beta}q^{\prime}_{\beta})_{V-A},\;\;
O_{10}=\frac{3}{2}(\bar{b}_{\alpha}X_{\beta})_{V-A}\sum_{q^{\prime}}e_{q^{\prime}}(\bar{q}^{\prime}_{\beta}q^{\prime}_{\alpha})_{V-A},
\end{eqnarray}
for the electro-weak penguin operators,
with the SU(3) color indices $\alpha$ and $\beta$ and the active quarks $q^{\prime}=(u,d,s,c,b)$ at the scale $m_{b}$.
The combinations   of the Wilson coefficients are defined as \cite{prd58094009}:
\begin{eqnarray}
&&a_{1}=C_{2}+C_{1}/3,\;\;\;\;\;\;a_{2}=C_{1}+C_{2}/3,\nonumber\\
&&a_{i}=C_{i}+C_{i\pm 1}/3,\;\;\;\;\; i=3,5,7,9\, / \,4,6,8,10.
\end{eqnarray}

  In the $B$ meson rest frame,   the light
final state mesons with large momenta are moving fast. The spectator quark in $B$ meson is soft in the initial state, while collinear
 in the final state. There must be a hard gluon to kick the soft spectator quark into collinear and energetic one. Thus the process
 is perturbatively calculable. The basic idea of PQCD approach is keeping the intrinsic transverse momentum $k_T$ of valence
  quarks in the hadrons. The end-point singularity in collinear factorization can be avoided. On the other hand, the double
  logarithms, which are caused by the additional energy scale introduced by the transverse momentum, can be resummed through
  the renormalization group equation to result in the Sudakov form factor. This form factor effectively suppresses the
  end-point contribution of distribution amplitude of mesons in the small transverse momentum region, which makes the calculation
   in the PQCD approach reliable and consistent.

In hadronic $B_{(s,c)}$ decays, there are several typical energy scales and expansions with respect to the ratios of the scales are usually
carried out, for example, the $W$ boson mass scale, the $b$ quark mass scale, and the factorization scale $\sqrt{\bar{\Lambda} m_B}$
 with $\bar{\Lambda}\equiv m_B-m_b$. We can perturbatively calculate the electroweak physics higher than W boson mass scale. The hard
  dynamics from $m_W$ scale to $m_b$ scale can be included in the so-called Wilson coefficients in eq.(\ref{eq:heff}) by using the
  renormalization group equation.
The physics between $M_{B}$ scale and the factorization scale can be calculated perturbatively and included in the so-called Hard
Kernel in the PQCD approach. The soft dynamics below the factorization scale is nonperturbative and described by the hadronic wave
functions of mesons, which is universal for all decay modes.  Finally, based on the factorization, the decay amplitude can be described as the following
convolution of the the Wilson coefficients $C(t)$, the hard scattering
kernel and the light-cone wave functions $\Phi_{h_{i},B}$ of mesons \cite{prd55and56},
\begin{eqnarray}
\mathcal
{A}\;\sim\;&&\int\,dx_{1}dx_{2}dx_{3}b_{1}db_{1}b_{2}db_{2}b_{3}db_{3}\nonumber\\
&&\times
Tr\left[C(t)\Phi_{B}(x_{1},b_{1})\Phi_{h_{2}}(x_{2},b_{2})\Phi_{h_{3}}(x_{3},b_{3})H(x_{i},b_{i},t)S_{t}(x_{i})e^{-S(t)}\right],
\end{eqnarray}
where $Tr$ denotes the trace over Dirac and colour indices, $b_{i}$ is the conjugate variable of quark's transverse
momentum $k_{iT}$, $x_{i}$ is the momentum fractions of valence quarks and $t$ is the largest energy scale in  the hard part
$H(x_{i},b_{i},t)$.   The jet function
$S_{t}(x_{i})$ smears the end-point singularities on $x_{i}$, which is from the threshold resummation of the double logarithms $\ln^2 x_i$
\cite{prd66094010}. The Sudakov form
factor  $e^{-S(t)}$ from the resummation of the double logarithms suppresses the soft dynamics effectively   i.e.
the long distance contributions in the large $b$ region
\cite{prd57443,lvepjc23275}.

For a tensor meson, the polarization tensor
$\epsilon_{\mu\nu}(\lambda)$ with helicity $\lambda$ can be expanded
through the polarization vectors $\epsilon^{\mu}(0)$ and
$\epsilon^{\mu}(\pm1)$ \cite{zheng1,zheng2}
\begin{eqnarray}
\epsilon^{\mu\nu}(\pm2)\,&\equiv&\,\epsilon(\pm1)^{\mu}\epsilon(\pm1)^{\nu},\nonumber\\
\epsilon^{\mu\nu}(\pm1)\,&\equiv&\,\sqrt{\frac{1}{2}}\left[\epsilon(\pm1)^{\mu}\epsilon(0)^{\nu}\,
+\,\epsilon(0)^{\mu}\epsilon(\pm1)^{\nu}\right],\nonumber\\
\epsilon^{\mu\nu}(0)\,&\equiv&\,\sqrt{\frac{1}{6}}\left[\epsilon(+1)^{\mu}\epsilon(-1)^{\nu}\,
+\,\epsilon(-1)^{\mu}\epsilon(+1)^{\nu}\right]\,+\,\sqrt{\frac{2}{3}}\epsilon(0)^{\mu}\epsilon(0)^{\nu}.
\end{eqnarray}
 In order to calculate
conveniently, a new polarization vector $\epsilon_{T}$ is defined as \cite{wwprd83014008}
\begin{eqnarray}
\epsilon_{T\mu}=\frac{1}{m_{B}}\epsilon_{\mu\nu}(h)P_{B}^{\nu},
\end{eqnarray}
with
\begin{eqnarray}
&&\epsilon_{T\mu}(\pm2)=0,\nonumber\\
&&\epsilon_{T\mu}(\pm1)=\frac{1}{m_{B}}\frac{1}{\sqrt{2}}\,\left(\epsilon(0)\cdot
P_{B}\right)\,\epsilon_{\mu}(\pm1),\nonumber\\
&&\epsilon_{T\mu}(0)=\frac{1}{m_{B}}\sqrt{\frac{2}{3}}\,\left(\epsilon(0)\cdot
P_{B}\right)\,\epsilon_{\mu}(0). \label{vector}
\end{eqnarray}
The $\pm 2$ polarizations do not contribute, which is
consistent with the angular momentum conservation argument in $B$
decays. The $\epsilon_{T}$ is similar with the $\epsilon$ of vector state,
regardless of the related constants \cite{wwprd83014008}. After this simplification,
 the wave function for a generic
tensor meson are defined by \cite{wwprd83014008}
\begin{eqnarray}
&&\Phi_{T}^{L}\,=\,\frac{1}{\sqrt{6}}\left[m_{T}\makebox[0pt][l]{/}\epsilon_{\bullet
L}^{*}\phi_{T}(x)\,+\,\makebox[0pt][l]{/}\epsilon_{\bullet
L}^{*}\makebox[-1.5pt][l]{/}P\phi_{T}^{t}(x)+m_{T}^{2}\frac{\epsilon_{\bullet}\cdot
v}{P\cdot v}\phi_{T}^{s}(x)\right]\nonumber\\
&&\Phi_{T}^{\perp}\,=\,\frac{1}{\sqrt{6}}\left[m_{T}\makebox[0pt][l]{/}\epsilon_{\bullet
\perp}^{*}\phi_{T}^{v}(x)\,+\,\makebox[0pt][l]{/}\epsilon_{\bullet
\perp}^{*}\makebox[-1.5pt][l]{/}P\phi_{T}^{T}(x)\,+\,m_{T}i\epsilon_{\mu\nu\rho\sigma}\gamma_{5}\gamma^{\mu}\epsilon_{\bullet
\perp}^{* \nu}n^{\rho}v^{\sigma}\phi_{T}^{a}(x)\right],
\end{eqnarray}
with the
vector
$\epsilon_{\bullet\mu}\,\equiv\,\frac{\epsilon_{\mu\nu}v^{\nu}}{P\cdot\,
v}$ related to the polarization tensor. The twist-2 and twist-3
distribution amplitudes are given by
\cite{wwprd83014008,zheng1,zheng2}
\begin{eqnarray}
&&\phi_{T}(x)\,=\,\frac{f_{T}}{2\sqrt{2N_{c}}}\phi_{\|}(x),\;\phi_{T}^{t}\,=\,\frac{f_{T}^{\perp}}{2\sqrt{2N_{c}}}h_{\|}^{(t)}(x),
\nonumber\\
&&\phi_{T}^{s}(x)\,=\,\frac{f_{T}^{\perp}}{4\sqrt{2N_{c}}}\frac{d}{dx}h_{\|}^{(s)}(x),\;
\phi_{T}^{T}(x)\,=\,\frac{f_{T}^{\perp}}{2\sqrt{2N_{c}}}\phi_{\perp}(x),\nonumber\\
&&\phi_{T}^{v}(x)\,=\,\frac{f_{T}}{2\sqrt{2N_{C}}}g_{\perp}^{(v)}(x),\;\phi_{T}^{a}(x)\,=
\,\frac{f_{T}}{8\sqrt{2N_{c}}}\frac{d}{dx}g_{\perp}^{(a)}(x),
\end{eqnarray}
with the form
\begin{eqnarray}
&&\phi_{\|,\perp}(x)\,=\,30x(1-x)(2x-1),\nonumber\\
&&h_{\|}^{(t)}(x)\,=\,\frac{15}{2}(2x-1)(1-6x+6x^{2}),\;h_{\|}^{(s)}(x)\,=\,15x(1-x)(2x-1),\nonumber\\
&&g_{\perp}^{(a)}(x)\,=\,20x(1-x)(2x-1),\;\;g_{\perp}^{(v)}(x)\,=\,5(2x-1)^{3}.
\end{eqnarray}
As   mentioned before, based on the Bose statistics, the above light-cone distribution
 amplitudes of the tensor meson are antisymmetric under the
 interchange of momentum fractions of the quark and anti-quark in
 the SU(3) limit (i.e. $x\leftrightarrow 1-x$) \cite{zheng1,zheng2}, which is consistent with the
 fact that  $<0\mid j^{\mu}\mid T>\,=\,0$, where $j^{\mu}$
is the $(V\pm A)$ or $(S\pm P)$ current.

All other meson wave functions $B_{u,d,s,c}$  and $\pi$, K etc. are adopted the same as in other pQCD papers\cite{annihilation1,annihilation2,prd78014018,pqcd2,pqcd3}, since we have argued that they should be universal for all decay channels.

\section{RESULTS AND DISCUSSIONS}

The traditional emission diagrams with a tensor meson emitted are prohibited in naive factorization, because a tensor meson can not be
produced from the local (V$\,\pm\,$A) or tensor currents. Thus, the predictions can
not be given within the naive factorization framework. The hard scattering contributions and annihilation type contributions
 are important to explain the large branching ratios.
When the emitted meson is the traditional light meson, the contributions from two hard scattering emission
  diagrams cancel with each other due to the symmetry.
 however, the symmetry is destroyed, when the tensor meson is emitted, because the distribution amplitudes
  of tensor meson are anti-symmetric. The two diagrams strengthen with each other. Thus,
  the hard scattering contributions are no longer negligible but sizable for the color suppressed modes with
  the enhancement of Wilson coefficients.

  \subsection{$B\to PT$ decays in PQCD approach}

\begin{table}
\centering
 \caption{The PQCD predictions of CP-averaged branching
ratios (in units of $10^{-6}$) for some $B\rightarrow PT$ decays, with the errors from hard parameters, NLO corrections and power corrections,  together with Isgur-Scora-Grinstein-Wise II (ISGW2)
model \cite{prd67014002} and QCDF results \cite{zheng2}. The
experimental data are from Ref.\cite{jpg37075021}.}
\begin{tabular}[t]{l!{\;\;\;\;}c!{\;\;\;}c!{\;\;\;\;\;\;\;}c!{\;\;\;\;\;\;\;}c!{\;\;\;\;\;\;\;}r}
\hline\hline \vspace{0.3cm}
 \multirow{1}{*}{Decay Modes} &\multirow{2}{*}{class}& \multirow{2}{*}{This Work} &\multirow{2}{*}{ISGW2 [7]}&\multirow{2}{*}{QCDF [14]} &\multirow{2}{*}{Expt.}\\
\hline \vspace{0.4cm}
\multirow{1}{*}{$B^{+}\rightarrow K_{2}^{*0}\pi^{+}$}& \multirow{1}{*}{PA} & \multirow{1}{*}{$0.9^{\;+0.2\;+0.2\;+0.3}_{\;-0.2\;-0.2\;-0.2}$} &\multirow{1}{*}{...}&\multirow{1}{*}{$3.1_{-3.1}^{+8.3}$}  &  \multirow{1}{*}{$5.6_{-1.4}^{+2.2}$}\\
\vspace{0.1cm}
$B^{+}\rightarrow f_{2}K^{+}$&T,PA,P&$12^{\;+3\;+3\;+3}_{\;-2\;-3\;-3}$&0.34&$3.8_{-3.0}^{+7.8}$&$1.06_{-0.29}^{+0.28}$\\
\vspace{0.1cm}
$B^{+}\rightarrow f_2^{\prime}K^{+}$&P,PA&$3.8^{\;+0.4\;+0.9\;+1.0}_{\;-0.4\;-0.8\;-0.8}$&0.004&$4.0_{-3.6}^{+7.4}$&$<7.7$\\
\vspace{0.1cm}
$B^{0}\rightarrow K_{2}^{*+}\pi^{-}$&PA&$1.0^{\;+0.2\;+0.2\;+0.3}_{\;-0.2\;-0.2\;-0.2}$&...&$3.3_{-3.2}^{+8.5}$&$<6.3$\\
\vspace{0.1cm}
$B^{0}\rightarrow f_{2}^{\prime}K^{0}$&P,PA&$3.7^{\;+0.3\;+0.7\;+0.9}_{\;-0.4\;-0.8\;-0.9}$&$0.00007$&$3.8_{-3.5}^{+7.3}$&...\\
\vspace{0.1cm}
$B^{0}\rightarrow a_{2}^{-}K^{+}$&T,PA&$5.0^{\;+1.6\;+1.4\;+1.3}_{\;-1.4\;-1.1\;-1.0}$&0.58&$9.7_{-8.1}^{+17.2}$&...\\
\vspace{0.1cm}
$B^{+}\rightarrow a_{2}^{0}\pi^{+}$&T,C &$29^{\;+13\;+14\;+3}_{\;-11\;-10\;-3}$&26.02&$30_{-12}^{+14}$ &...\\
\vspace{0.1cm}
$B^{+}\rightarrow a_{2}^{+}\pi^{0}$&T,C& $0.3^{\;+0.0\;+0.1\;+0.0}_{\;-0.0\;-0.1\;-0.0}$&0.01&$2.4_{-3.1}^{+4.9}$& ...\\
\vspace{0.1cm}
$B^{+}\rightarrow a_{2}^{+}\eta^{\prime}$&C,PA,P&$3.5^{\;+1.4\;+1.6\;+1.1}_{\;-1.0\;-1.1\;-0.8}$&13.1&$1.1_{-1.2}^{+4.7}$&...\\
\vspace{0.1cm}
$B^{+}\rightarrow f_{2}\pi^{+}$&T&$43^{\;+19\;+19\;+4}_{\;-15\;-14\;-4}$&28.74&$27_{-12}^{+14}$&$15.7_{-4.9}^{+6.9}$\\
\vspace{0.1cm}
$B^{+}\rightarrow f_{2}^{\prime}\pi^{+}$&T&$1.2^{\;+0.3\;+0.4\;+0.1}_{\;-0.2\;-0.3\;-0.1}$&0.37&$0.09_{-0.09}^{+0.24}$&...\\
\vspace{0.1cm}
$B^{0}\rightarrow a_{2}^{-}\pi^{+}$&T&$99^{\;+35\;+43\;+6}_{\;-30\;-32\;-10}$&48.82&$52_{-18}^{+18}$&$<3000$\\
\vspace{0.1cm}
$B^{0}\rightarrow a_{2}^{+}\pi^{-}$&T,PA&$2.7^{\;+0.5\;+0.8\;+0.4}_{\;-0.3\;-0.5\;-0.3}$&...&$2.1_{-1.7}^{+4.3}$&...\\
\vspace{0.1cm}
$B^{0}\rightarrow f_{2}\pi^{0}$&C&$2.8^{\;+0.7\;+0.7\;+0.6}_{\;-0.6\;-0.6\;-0.4}$&0.003&$1.5_{-1.4}^{+4.2}$&...\\
\vspace{0.1cm}
$B^{0}\rightarrow f_{2}^{\prime}\pi^{0}$&P&$0.2^{\;+0.0\;+0.1\;+0.0}_{\;-0.0\;-0.1\;-0.0}$&$4.0\times 10^{-5}$&$0.05_{-0.05}^{+0.12}$&...\\
 \hline\hline
\end{tabular}\label{tb:pt1}
\end{table}

 From  Table.\ref{tb:pt1}, one can see that
the $CP$-averaged branching ratios of $B\to PT$ decays  can reach  the order of $10^{-5}$, which is accessible at the current experiments. The
predicted branching ratios of penguin-dominated and color-suppressed
decays in PQCD are larger than those of naive factorization
\cite{prd67014002,jpg36095004,arxiv1010.3077}, but are close to the  QCDF predictions \cite{zheng2},
which is caused by the fact that the naive factorization can not evaluate the contributions from penguin operators well.
On the other hand, the $B$ to tensor form factor calculated in PQCD approach is larger than that used in QCDF \cite{zheng2}, thus
for the tree-dominated modes such as $a_{2}^{-}\pi^{+}$ and
$f_{2}\pi^{+}$, the predicted results are bigger than QCDF predictions \cite{zheng2}.
Although the $B^+ \to a_{2}^{0}\pi^{+}$ is also tree-dominant, the result is the same as the prediction of QCDF,
which is caused by the cancellation from hard scattering contributions with a tensor emitted.

For $B^{+}\rightarrow
K_{2}^{*0}\pi^{+}$ and $B^{0}\rightarrow K_{2}^{*+}\pi^{-}$ decays, the predictions in naive factorization approach \cite{prd67014002} are 0 indeed, because  the emitted meson is the tensor meson $K_2^*$.
In PQCD approach, the expected hard scattering contributions are suppressed by the small Wilson coefficients.
Thus the dominant contribution is from the chiral enhanced annihilation type contributions. While in QCDF, the dominant
contribution comes from hard scattering diagrams with quark loop corrections, which is next-to-leading order and not considered in this work.
 The branching ratios of  $a_{2}^{+}\pi^{-}$ and $a_{2}^{+}\pi^{0}$
modes are highly suppressed relative to $a_{2}^{-}\pi^{+}$ and
$a_{2}^{0}\pi^{+}$, respectively, because the expected dominant contribution of $B\rightarrow
a_{2}^{+}\pi^{0}(a_{2}^{+}\pi^{-})$ is the color favored tensor emission diagram, while the dominant contribution for the other two channels  is the color enhanced diagram with pion emission.
Many other smaller branching ratios for $B\to PT$ decays are listed in Tables of Ref.\cite{zoupt}.

 \begin{table}
\centering
 \caption{The PQCD predictions of  direct CP asymmetries($\%$) for $B\rightarrow PT$ decays, comparison with the QCDF results \cite{zheng2}. The
experimental data are from Ref.\cite{jpg37075021}.}
 \vspace{0.3cm}
\begin{tabular}{l!{\;\;\;\;\;\;\;\;\;}c!{\;\;\;\;\;\;\;\;\;}c!{\;\;\;\;\;\;\;\;\;}r}
\hline\hline\vspace{0.3cm}
 \multirow{1}{*}{Decay Modes} & \multirow{1}{*}{This Work} & \multirow{1}{*}{QCDF [14]} &\multirow{1}{*}{Expt.}\\
\hline \vspace{0.4cm}
\multirow{1}{*}{$B^{+}\rightarrow K_{2}^{*0}\pi^{+}$} & \multirow{1}{*}{$-5.5^{\;+0.3\;+2.6\;+1.6}_{\;-0.4\;-0.0\;-1.2}$} & \multirow{1}{*}{$1.6_{-1.8}^{+2.2}$}  & \multirow{1}{*}{$5^{+29}_{-24}$}\\
\vspace{0.1cm}
$B^{+}\rightarrow f_{2}K^{+}$&$-25^{\;+2\;+2\;+5}_{\;-1\;-3\;-6}$&$-39.5_{-25.5}^{+49.4}$&$-68.0_{-17}^{+19}$\\
\vspace{0.1cm}
$B^{+}\rightarrow K_{2}^{*+}\eta$&$-5.4^{\;+1.1\;+2.2\;+2.3}_{\;-0.6\;-2.0\;-1.3}$&$1.5_{-5.6}^{+7.4}$&$-45\pm30$\\
\vspace{0.1cm}
$B^{0}\rightarrow a_{2}^{-}K^{+}$&$-48^{\;+2\;+1\;+7}_{\;-2\;-0\;-10}$&$-21.5_{-35.0}^{+28.9}$&...\\
\vspace{0.1cm}
$B^{+}\rightarrow a_{2}^{+}\eta$    &$-91^{\;+8\;+10\;+12}_{\;-4\;\;-1\;\;\;-5}$&$27.6_{-127.6}^{+73.4}$&...\\
\vspace{0.1cm}
$B^{+}\rightarrow a_{2}^{+}\eta^{\prime}$&$-45^{\;+1\;+1\;+7}_{\;-1\;-0\;-9}$&$31.3_{-131.3}^{+61.3}$&...\\
\vspace{0.1cm}
$B^{+}\rightarrow f_{2}\pi^{+}$&$28^{\;+3\;+1\;+9}_{\;-3\;-1\;-7}$&$60.2_{-72.3}^{+27.1}$&$41\pm30$\\
\vspace{0.1cm}
$B^{0}\rightarrow f_{2}^{\prime}\eta$&$71^{\;+0\;+11\;+11}_{\;-3\;-15\;-12}$&$0.0$&...\\
\vspace{0.1cm}
$B^{0}\rightarrow f_{2}^{\prime}\eta^{\prime}$&$46^{\;+3\;+14\;+19}_{\;-7\;-12\;-19}$&$0.0$&...\\
\hline\hline
\end{tabular}\label{tb:pt2}
\end{table}

We also summarize the   direct $CP$ asymmetries for some $B\rightarrow PT$ decays in Table \ref{tb:pt2}.
The full results about $CP$ violation can be found in Ref.\cite{zoupt}.
Although some channels have very large direct CP asymmetries, they are difficult for experiments due to the small branching ratios.
 We recommend the experimenters to
search for the direct CP asymmetry in the channels like $B^+\to f_2 K^+$, $B^0
\to a_2^- K^+$, $B^+\to a_2^+\eta'$ and $B^+ \to f_2 \pi^+$, for
they have both large branching ratios and direct CP asymmetry parameters. In fact, there are already some experimental measurements
shown in Table~\ref{tb:pt2}. Although the error bars are
still large, we are happy to see that all these measured entries
have the same sign as our theoretical calculations. This may imply
that our approach gives the leading order strong phase in these channels.

Similar to the $\eta-\eta^{\prime}$ system, we have  taken
the $f_{2}-f_{2}^{\prime}$ mixing into account \cite{zoupt}. Although the mixing angle is small, the
interference between $f_{2}^q$ and $f_{2}^{s}$ can bring some
remarkable changes. For example, the branching ratio of
$B^{+}\rightarrow f_{2}^{\prime}\pi^{+}$ is  enhanced comparing with the
prediction of QCDF \cite{zheng2} without the mixing. Although suppressed by the small mixing angle,
the color-allowed contribution from $f_2^q$ term, which is not considered in QCDF \cite{zheng2},
is at the same order as that of $f_2^s$ term. The enhancement from $f_{2}^{q}$ term makes the
branching ratio larger than the prediction without taking
the mixing into account. The mixing can also bring remarkable change
to direct CP asymmetry.
Without the mixing, the direct $CP$ asymmetry of $B\rightarrow f_{2}^{\prime}\eta^{(\prime)}$ decay should be 0,
since there are no contributions from penguin operators.
After taking the mixing into account, the direct CP
asymmetries are    quite large shown in Table \ref{tb:pt2},
since these decays get penguin contributions from $f_{2}^{q}$.

  \subsection{$B_{s}\to PT$ decays in PQCD approach}

We have also studied the two-body hadronic $B_s \to PT$ decays in the PQCD approach
and given the predictions about branching ratios and $CP$ observables \cite{bspt}. Like $K^0-\bar{K}^0$,
$D^0-\bar{D}^0$ and $B^0-\bar{B}^0$ systems, the $B^0_s-\bar{B}^0_s$ also mixes  through
the weak interaction.  Since the mass difference $\Delta M$ between the mass eigenstates is much larger than
the decay width $\Gamma$ of the $B_s$ meson, the $B^0_s-\bar{B}^0_s$ oscillate very frequent. As a result, the measurements,
such as time-dependent CP-violation parameters, are very difficult for super $B$ factories, while are feasible
in LHCb experiments.

For $\Delta S=0$ decays, the $B_s^0$ ($\bar{B}_s^0$) meson decays to the final
state $f$ ($\bar{f}$), but not to $\bar{f}$ ($f$) with $f\neq\bar{f}$.
One can find  in Table \ref{tb:bspt1} that only the $B_s^0\to\pi^+K_2^{*-}$ decay has
a sizable branching ratio due to the color enhanced emission amplitude $T$. Because the
factorizable emission contributions are suppressed, those
color-suppressed modes, such as $B_s^0\to\bar{K}^0a_2^0$, $\bar{K}^0f_2$ and
$\bar{K}^0f_2'$, are similar with their $PV$ partners \cite{prd76074018}.
While for the color-favored
$B_s^0\to K^-a_2^+$ decay, the branching ratio $1.50\times10^{-7}$ is much smaller than the
$B_s^0\to K^-\rho^+$ one, $1.78\times10^{-5}$, because the factorizable emission amplitude, which is dominant in $B_s^0\to K^-\rho^+$, is
forbidden in naive factorization with a tensor meson emission.
Combing those predictions with the $B\to PT$ ones,
we  find large U-spin asymmetries in some decay modes, such as $B_s^0\to\pi^+K_2^{*-}$ and $B_s^0\to K^-a_2^+$ \cite{bspt}.

\begin{table}
  \centering
  \caption{Some results about branching ratios ($10^{-7}$) and direct CP asymmetries of
  the $\Delta$S=0 $B_s^0\to PT$ decays.}
  \begin{tabular}[t]{cccc}\hline\hline
  Modes & Class & $Br (10^{-7})$ & Direct $A_{CP}$ ($\%$)\\\hline
  $B_s^0\to \pi^+K_2^{*-}$ & $T$ & $90^{+40+4}_{-32-6}$ & $13^{+2+2}_{-2-2}$\\
  $B_s^0\to \bar{K}^0a_2^0$ & $C$,$PA$ & $2.0^{+0.4+0.2}_{-0.3-0.3}$ & $38^{+7+6}_{-10-7}$ \\
  $B_s^0\to \bar{K}^0f_2$ & $C$,$PA$ & $3.4^{+0.7+0.7}_{-0.6-0.7}$ & $-24^{+5+3}_{-6-5}$ \\
  $B_s^0\to K^-a_2^+$ & $T$,$PA$ & $1.5^{+0.3+0.4}_{-0.2-0.3}$ & $39^{+8+1}_{-1-4}$ \\
  \hline\hline
  \end{tabular}\label{tb:bspt1}
\end{table}

For some $\Delta S=1$ $B_s^0$ ($\bar{B}_s^0$) meson decays, whose final states are CP eigenstates,
i.e. $f=\bar{f}$, the results are summarized in Table~\ref{tb:bspt2}. One can find that
$B_s^0\to\eta' a_2^0(f_2,f_2')$ have branching ratios larger than those of the $B_s^0\to\eta a_2^0(f_2,f_2')$ modes,
because the dominant $\bar{s}s$ constituent is suppressed for decays involving $\eta$ than $\eta'$.
On the other hand, the $\Delta I=1$ modes, like
$B_s^0\to\eta a_2^0$ and $\eta' a_2^0$, are highly suppressed, compared
to the corresponding $\Delta I=0$ modes, $B_s^0\to\eta f_2$ and
$\eta' f_2$, because the dominant contributions from different penguin
operators are destructive in former modes, but constructive in latter decays.
Contrary to the $\Delta S=0$ decays, the direct CP violation
$C_f$'s are tiny, because the contributions from tree operators are too small.

\begin{table}
  \centering
  \caption{Some results of branching ratios ($10^{-7}$) and CP observables for
  the $\Delta S=1$ $B_s^0\to PT$ decays.}\label{tb:bspt2}
  \begin{tabular}[t]{ccccccc}\hline\hline
  Modes & Class & $Br$ & $C_f$ & $D_f$ & $S_f$ & $A_{CP}$ ($\%$)\\\hline
  $\eta a_2^0$ & $C$,$A$ & $0.047^{+0.013+0.010}_{-0.010-0.012}$ & $0.02^{+0.01+0.01}_{-0.02-0.06}$ & $0.40^{+0.01+0.06}_{-0.01-0.04}$ & $0.92^{+0.01+0.02}_{-0.01-0.03}$ & -3.6\\
  $\eta f_2$ & $PC$ & $9.8^{+2.7+3.2}_{-2.2-2.6}$ & $-0.014^{+0.003+0.008}_{-0.008-0.010}$ & $-0.995^{+0.001+0.002}_{-0.001-0}$ & $-0.098^{+0.007+0.004}_{-0.007-0.020}$ & 0.30\\
  $\eta f_2'$ & $PA$ & $96^{+20+36}_{-19-30}$ & $0.022^{+0.004+0.003}_{-0.004-0.003}$ & $-1.000^{+0+0}_{-0-0}$ & $0.024^{+0.004+0.003}_{-0.004-0.005}$ & -0.10\\
  $\eta' a_2^0$ & $C$,$A$ & $0.13^{+0.03+0.03}_{-0.03-0.03}$ & $0.03^{+0.01+0.02}_{-0.01-0.01}$ & $0.28^{+0.03+0.04}_{-0-0.03}$ & $0.96^{+0+0.01}_{-0.01-0.01}$ & -3.7\\
  $\eta' f_2$ & $PC$ & $30^{+7+11}_{-7-10}$ & $-0.005^{+0+0.002}_{-0.012-0.010}$ & $-0.994^{+0.001+0.001}_{-0.001-0.001}$ & $-0.104^{+0.011+0.006}_{-0.006-0.006}$ &0.40\\
  $\eta' f_2'$ & $PA$,$PT$ & $245^{+69+99}_{-59-84}$ & $-0.007^{+0.004+0}_{-0.003-0.001}$ & $-1.000^{+0+0}_{-0-0}$ & $-0.009^{+0.006+0.004}_{-0.002-0.001}$ & 0.030\\
  \hline\hline
  \end{tabular}
\end{table}

\begin{table}[!htbh]
  \centering
  \caption{Branching ratios (in units of $10^{-7}$) and CP observables for the rest $\Delta S=1$ decays.}\label{tb:bspt3}
  \begin{tabular}[t]{ccccc}\hline\hline
  Modes & $C_f$ & $D_f$ & $S_f$ & $C_{\bar{f}}$  \\\hline
  $\pi^+a_2^-$  & $-0.15^{+0.01+0.02}_{-0.04-0.05}$ & $-0.98^{+0+0.01}_{-0-0.01}$ & $-0.10^{+0.07+0.05}_{-0.01-0.01}$  & $-0.05^{+0.07+0.07}_{-0.02-0.01}$ \\
  $K^+K_2^{*-}$  & $0.49^{+0.07+0.02}_{-0.06-0.01}$ & $-0.85^{+0.04+0}_{-0.03-0}$ & $-0.18^{+0.02+0.03}_{-0.04-0.05}$  & $0.03^{+0.11+0.09}_{-0.08-0.13}$ \\
  $K^0\bar{K}_2^{*0}$  & $0.24^{+0.08+0.03}_{-0.06-0.05}$ & $-0.91^{+0.03+0.02}_{-0.02-0.02}$ & $-0.34^{+0.03+0.04}_{-0.03-0.03}$ & $0.24^{+0.08+0.03}_{-0.06-0.05}$ \\\hline
  &$D_{\bar{f}}$ & $S_{\bar{f}}$ & $Br$ & $A_{CP} ($\%$)$\\\hline
  $\pi^+a_2^-$& $-0.98^{+0.01+0.01}_{-0.01-0.01}$ & $0.18^{+0.04+0.04}_{-0.02-0.03}$ & $1.8^{+0.4+0.6}_{-0.2-0.8}$ & $13^{+3+5}_{-5-5}$\\
  $K^+K_2^{*-}$& $-0.71^{+0.09+0.03}_{-0.06-0.02}$ & $-0.70^{+0.07+0.03}_{-0.07-0.03}$ & $86^{+20+28}_{-16-24}$ & $-28^{+2+5}_{-3-6}$\\
  $K^0\bar{K}_2^{*0}$& $-0.91^{+0.03+0.02}_{-0.02-0.02}$ & $-0.34^{+0.03+0.04}_{-0.03-0.03}$ & $70^{+14+24}_{-12-20}$ & 0\\
  \hline\hline
  \end{tabular}
\end{table}

There exist more complicated $\Delta S=1$ modes, in which either a $B_s^0$ or
$\bar{B}_s^0$ meson can decay into $f$ or $\bar{f}$ with $f\neq\bar{f}$. The results are summarized in Table \ref{tb:bspt3}.
One can find that, for $B_s^0\to\bar{K}^0K_2^{*0}$ , the $f$-related CP observables are the
same as the $\bar{f}$-related ones. What's more, $A_{CP}$ is exactly zero due to the shortage of tree
contributions. It is then straightforward to arrive at
$\lambda_f=\bar{\lambda}_{\bar{f}}$, and thus
$C(D,S)_f=C(D,S)_{\bar{f}}$ and $A_{CP}=0$.

\subsection{$B_{(s)} \to D^{(*)}T,\bar{D}^{(*)}T$ decays in PQCD approach}
\begin{table}[htb]
\centering
 \caption{Branching ratios of some $B_{(s)}\rightarrow DT$ decays in the PQCD approach
 together with results from Isgur-Scora-Grinstein-Wise (ISGW) II model \cite{arxiv1010.3077,prd67014011} (unit:$10^{-7}$).}

\begin{tabular}[t]{l!{\;\;\;\;}c!{\;\;\;\;}c!{\;\;\;\;}c!{\;\;\;\;}c}
\hline\hline

 \multirow{2}{*}{Decay Modes}& \multirow{2}{*}{Class} & \multirow{2}{*}{This Work} &
 \multirow{2}{*}{SDV[9]}
 & \multirow{2}{*}{KLO[10]}\\
 &&&&\\
 \hline
\vspace{0.13cm}
\multirow{1}{*}{$B^{0}\rightarrow D^{0}f_{2}$}& \multirow{1}{*}{C}& \multirow{1}{*}{$2.1_{-0.7\,-0.3\,-0.2}^{+0.8\,+0.3\,+0.3}$} & \multirow{1}{*}{0.36} &\multirow{1}{*}{...}\\
\vspace{0.13cm}
\multirow{1}{*}{$B^{0}\rightarrow D^{0}f_{2}^{\prime}$} & \multirow{1}{*}{C} & \multirow{1}{*}{$0.038_{-0.013\,-0.006\,-0.005}^{+0.015\,+0.006\,+0.005}$} & \multirow{1}{*}{$0.0071$} & \multirow{1}{*}{...}\\
\vspace{0.13cm}
\multirow{1}{*}{$B^{+}\rightarrow D^{0}K_{2}^{*+}$} & \multirow{1}{*}{C} & \multirow{1}{*}{$37_{-12\,-8\,-5}^{+14\,+7\,+5}$} & \multirow{1}{*}{13} & \multirow{1}{*}{12}\\
\vspace{0.13cm}
\multirow{1}{*}{$B_{s}\rightarrow D^{0}\bar{K}_{2}^{*0}$} & \multirow{1}{*}{C}&\multirow{1}{*}{$1.4_{-0.6\,-0.2\,-0.2}^{+0.7\,+0.2\,+0.2}$} & \multirow{1}{*}{0.46} & \multirow{1}{*}{...}\\
\vspace{0.13cm}
\multirow{1}{*}{$B^{0}\rightarrow D^{+}a_{2}^{-}$} & \multirow{1}{*}{T} & \multirow{1}{*}{$15_{-6\,-3\,-2}^{+8\,+2\,+2}$} & \multirow{1}{*}{12} & \multirow{1}{*}{...}\\
\vspace{0.13cm}
\multirow{1}{*}{$B^{+}\rightarrow D^{+}a_{2}^{0}$} & \multirow{1}{*}{T} & \multirow{1}{*}{$9.4_{-3.4\,-1.6\,-1.1}^{+4.6\,+1.2\,+1.2}$} & \multirow{1}{*}{6.5} & \multirow{1}{*}{...}\\
\vspace{0.13cm}
\multirow{1}{*}{$B_{s}\rightarrow D^{+}K_{2}^{*-}$} & \multirow{1}{*}{T} & \multirow{1}{*}{$11_{-5\,-1\,-1}^{+6\,+1\,+1}$} & \multirow{1}{*}{8.3} & \multirow{1}{*}{...}\\
\vspace{0.13cm}
\multirow{1}{*}{$B^{0}\rightarrow D_{s}^{+}K_{2}^{*-}$} & \multirow{1}{*}{E} & \multirow{1}{*}{$0.61_{-0.14\,-0.16\,-0.07}^{+0.15\,+0.12\,+0.08}$} & \multirow{1}{*}{...} & \multirow{1}{*}{...}\\
\vspace{0.13cm}
\multirow{1}{*}{$B^{+}\rightarrow D^{+}K_{2}^{*0}$} & \multirow{1}{*}{A} & \multirow{1}{*}{$5.3_{-1.7\,-0.7\,-0.7}^{+1.8\,+0.7\,+0.7}$} & \multirow{1}{*}{...} & \multirow{1}{*}{...}\\
\vspace{0.3cm}
\multirow{1}{*}{$B_{s}\rightarrow D^{0}a_{2}^{0}$} & \multirow{1}{*}{E} & \multirow{1}{*}{$3.9_{-1.2\,-1.0\,-0.5}^{+1.4\,+0.7\,+0.5}$} & \multirow{1}{*}{...} & \multirow{1}{*}{...}\\
 \hline\hline
\end{tabular}\label{tb:dt1}
\end{table}

 There are two categories of decays with one charmed meson in the final states. The $B_{(s)}\rightarrow \bar{D}^{(*)}T$ decays governed by the $\bar b \to \bar c$ transition, while the $B_{(s)}\rightarrow D^{(*)} T$ decays governed by the $\bar b \to \bar u $ transition. Clearly, there is  a large enhancement of CKM matrix elements $|V_{cb}/V_{ub}|^2$ for the the former kinds of decays, especially for those
without a strange quark in the four-quark operators.
 One can find that the branching ratios of $B_{(s)}\rightarrow \bar{D}^{(*)}T$ decays are larger than those of $B_{(s)}\rightarrow D^{(*)} T$
decays shown in Tables~\ref{tb:dt1},\ref{tb:dt2} and Tables~\ref{tb:dt3},\ref{tb:dt4}, respectively. For most of
the $B_{(s)}\rightarrow D^{(*)} T$ decays, the branching ratios are at the
order $10^{-6}$ or $10^{-7}$; while for the $B_{(s)}\rightarrow
\bar{D}^{(*)}T$ decays, the branching ratios are at the order
$10^{-4}$ or $10^{-5}$. With a charm quark in the final state, all these decays are governed by tree level current-current operators, without penguin operator contribution.
Since the direct CP asymmetry is proportional to
the interference between two different contributions.
  all these decays have no direct $CP$ asymmetries.

\begin{table}
\centering
 \caption{Branching ratios (unit:$10^{-7}$) and the percentage of transverse polarizations $R_{T}$(unit:$\%$) of some $B_{(s)}\rightarrow D^{*}T$ decays in the PQCD approach
 together with results from ISGW II model \cite{arxiv1010.3077,prd67014011}.}
\begin{tabular}[t]{lccccc}
\hline\hline
 \vspace{0.2cm}
 \multirow{3}{*}{Decay Modes}& \multirow{3}{*}{Class} &
 \multicolumn{3}{c}{ \multirow{1}{*}{Branching Ratio($10^{-7}$)}}& \multirow{3}{*}{$R_{T}(\%)$}\\
\cline{3-5}
&&\multirow{2}{*}{This Work}&\multirow{2}{*}{SDV[9]}& \multirow{2}{*}{KLO[10]}&\\
&&&&&\\
 \hline
 \vspace{0.13cm}
\multirow{1}{*}{$B^{0}\rightarrow D^{*0}f_{2}$}& \multirow{1}{*}{C}&
\multirow{1}{*}{$2.7_{-1.0\,-0.3\,-0.3}^{+1.2\,+0.4\,+0.4}$} & \multirow{1}{*}{0.53} &
\multirow{1}{*}{...}& \multirow{1}{*}{$26_{-4\,-1}^{+4\,+1}$}\\
\vspace{0.13cm}
\multirow{1}{*}{$B^{+}\rightarrow D^{*0}K_{2}^{*+}$} & \multirow{1}{*}{C} &
 \multirow{1}{*}{$72_{-24\,-9\,-9}^{+28\,+12\,+10}$} & \multirow{1}{*}{21} &
  \multirow{1}{*}{19}&\multirow{1}{*}{$35_{-4\,-1}^{+4\,+1}$}\\
\vspace{0.13cm}
\multirow{1}{*}{$B_{s}\rightarrow D^{*0}\bar{K}_{2}^{*0}$} & \multirow{1}{*}{C}&
\multirow{1}{*}{$2.1_{-0.8\,-0.3\,-0.2}^{+1.0\,+0.3\,+0.3}$} & \multirow{1}{*}{0.7} &
\multirow{1}{*}{...}&\multirow{1}{*}{$21_{-4\,-1}^{+3\,+1}$}\\
\vspace{0.13cm}
\multirow{1}{*}{$B^{+}\rightarrow D^{*+}K_{2}^{*0}$} & \multirow{1}{*}{A} &
\multirow{1}{*}{$18_{-5\,-2\,-3}^{+5\,+0\,+2}$} & \multirow{1}{*}{...} &
 \multirow{1}{*}{...}& \multirow{1}{*}{$82_{-3\,-3}^{+2\,+4}$}\\
\vspace{0.13cm}
\multirow{1}{*}{$B^{+}\rightarrow D_{s}^{*+}f_{2}^{\prime}$}& \multirow{1}{*}{A} &
 \multirow{1}{*}{$22_{-6\,-2\,-3}^{+7\,+1\,+3}$} & \multirow{1}{*}{4.0} &
 \multirow{1}{*}{2.0}& \multirow{1}{*}{$83_{-5\,-2}^{+5\,+2}$}\\
\vspace{0.13cm}
\multirow{1}{*}{$B^{+}\rightarrow D_{s}^{*+}\bar{K}_{2}^{*0}$} & \multirow{1}{*}{A} &
\multirow{1}{*}{$1.3_{-0.3\,-0.2\,-0.2}^{+0.4\,+0.1\,+0.2}$} & \multirow{1}{*}{...} & \multirow{1}{*}{...}&\multirow{1}{*}{$81_{-2\,-3}^{+2\,+4}$}\\
\vspace{0.13cm}
\multirow{1}{*}{$B^{0}\rightarrow D_{s}^{*+}K_{2}^{*-}$} & \multirow{1}{*}{E} &
\multirow{1}{*}{$0.57_{-0.13\,-0.11\,-0.07}^{+0.13\,+0.12\,+0.07}$} & \multirow{1}{*}{...} &
\multirow{1}{*}{...}&\multirow{1}{*}{$12_{-2\,-3}^{+2\,+3}$}\\
\vspace{0.3cm}
\multirow{1}{*}{$B_{s}\rightarrow D^{*+}a_{2}^{-}$} & \multirow{1}{*}{E} & \multirow{1}{*}{$5.4_{-1.6\,-1.3\,-0.7}^{+1.8\,+1.4\,+0.7}$} & \multirow{1}{*}{...} & \multirow{1}{*}{...}&\multirow{1}{*}{$21_{-3\,-4}^{+3\,+5}$}\\
 \hline\hline
\end{tabular}\label{tb:dt2}
\end{table}

Due to the symmetry, the contributions from two hard scattering emission diagrams
cancel with each other in the charmless $B\to PP$, $PV$ decays,   thus the hard scattering emission diagrams are suppressed heavily.
However, when the emitted meson is the
$D(\bar{D})$ or tensor meson, the cancellation is weakened or removed. The symmetry of those two
hard scattering emission diagrams is broken by the big difference between $c(\bar{c})$ quark and the light
quark in the heavy $D(\bar{D})$ meson.  As stated above, when the tensor meson is
emitted, the contributions from two hard scattering diagrams
 strengthen with each other, because the wave function of
tensor meson is antisymmetric under the interchange of the momentum
fractions of the quark and antiquark.
As a result, for those color suppressed decay modes,
the hard scattering contribution plays the crucial role in the decay amplitude, since the factorizable contributions are suppressed by the small Wilson coefficient $a_2$. Therefore, the predicted branching ratios
in the PQCD approach are larger than those of naive factorization \cite{arxiv1010.3077,prd67014011}. Taking
$B^{0}\rightarrow \bar{D}^{0}f_{2}$ as example, the PQCD prediction
$\mathcal {B}(B^{0}\rightarrow
\bar{D}^{0}f_{2})\,=\,9.46\times10^{-5}$, which is larger than other
approaches,  agrees better with the experimental data
$(12\pm4)\times10^{-5}$ \cite{jpg37075021}. For those decays with a tensor meson emitted,
 since the factorizable contributions are prohibited,
 the naive factorization can not give predictions \cite{arxiv1010.3077,prd67014011}. But these
decays can get contributions from hard scattering and annihilation
type diagrams.  The PQCD approach can calculate these contributions and give the predictions for the first time in
table \ref{tb:dt1}-\ref{tb:dt4}. Our results show that the contributions from
annihilation diagrams are even at the same order as the emission
diagrams in some decay modes. The interference between contributions from emission diagram and
contributions from annihilation diagrams can explain why $\mathcal
{B}(B^{0}\rightarrow \bar{D}^{(*)0}a_{2}^{0})\,>\,\mathcal
{B}(B^{0}\rightarrow \bar{D}^{(*)0}f_{2})$, which is similar to $\mathcal {B}(B^{0}\rightarrow
\bar{D}^{(*)0}\rho^{0})\,>\,\mathcal {B}(B^{0}\rightarrow
\bar{D}^{(*)0}\omega)$ \cite{fd1}.

\begin{table}
\centering
 \caption{Branching ratios of some $B_{(s)}\rightarrow \bar{D}T$ decays calculated in the PQCD approach
 together with results from ISGW II model \cite{arxiv1010.3077,prd67014011} (unit:$10^{-5}$).}
\begin{tabular}[t]{lcccc}
\hline\hline

 \multirow{2}{*}{Decay Modes}& \multirow{2}{*}{Class} & \multirow{2}{*}{This Work} &
 \multirow{2}{*}{SDV[9]}
 & \multirow{2}{*}{KLO[10]}\\
 &&&&\\
 \hline
 \vspace{0.13cm}
\multirow{1}{*}{$B^{0}\rightarrow \bar{D}^{0}a_{2}^{0}$}&\multirow{1}{*}{C} &
\multirow{1}{*}{$12_{-3\,-3\,-0}^{+3\,+3\,+1}$} &\multirow{1}{*}{8.2}&\multirow{1}{*}{4.8}\\
\vspace{0.13cm}
\multirow{1}{*}{$B^{0}\rightarrow \bar{D}^{0}f_{2}$}& \multirow{1}{*}{C}&
\multirow{1}{*}{$9.5_{-2.3\,-3.7\,-0.3}^{+2.5\,+3.6\,+0.5}$} & \multirow{1}{*}{8.8} & \multirow{1}{*}{5.3}\\
\vspace{0.13cm}
\multirow{1}{*}{$B^{+}\rightarrow \bar{D}^{0}a_{2}^{+}$} & \multirow{1}{*}{T,C} &
\multirow{1}{*}{$42_{-13\,-14\,-1}^{+17\,+13\,+2}$} & \multirow{1}{*}{18} & \multirow{1}{*}{10}\\
\vspace{0.13cm}
\multirow{1}{*}{$B_{s}\rightarrow \bar{D}^{0}\bar{K}_{2}^{*0}$} & \multirow{1}{*}{C} &
 \multirow{1}{*}{$20_{-6\,-5\,-1}^{+8\,+4\,+1}$} & \multirow{1}{*}{11} & \multirow{1}{*}{...}\\
\vspace{0.13cm}
\multirow{1}{*}{$B^{0}\rightarrow D^{-}a_{2}^{+}$} & \multirow{1}{*}{T} &
 \multirow{1}{*}{$40_{-13\,-12\,-1}^{+15\,+13\,+2}$} & \multirow{1}{*}{...} & \multirow{1}{*}{...}\\
\vspace{0.13cm}
\multirow{1}{*}{$B^{0}\rightarrow D^{-}K_{2}^{*+}$} & \multirow{1}{*}{T} &
\multirow{1}{*}{$1.2_{-0.4\,-0.5\,-0.1}^{+0.5\,+0.5\,+0.1}$} & \multirow{1}{*}{...} & \multirow{1}{*}{...}\\
\vspace{0.13cm}
\multirow{1}{*}{$B_{s}\rightarrow D_{s}^{-}a_{2}^{+}$} & \multirow{1}{*}{T} &
\multirow{1}{*}{$11_{-4\,-5\,-0}^{+6\,+6\,+1}$} & \multirow{1}{*}{...} & \multirow{1}{*}{...}\\
\vspace{0.13cm}
\multirow{1}{*}{$B^{0}\rightarrow D_{s}^{-}K_{2}^{*+}$} & \multirow{1}{*}{E} &
\multirow{1}{*}{$6.1_{-1.7\,-1.0\,-0.2}^{+1.7\,+0.4\,+0.3}$} & \multirow{1}{*}{...} & \multirow{1}{*}{...}\\
\vspace{0.3cm}
\multirow{1}{*}{$B_{s}\rightarrow D^{-}a_{2}^{+}$} & \multirow{1}{*}{E} &
\multirow{1}{*}{$0.23_{-0.08\,-0.04\,-0.01}^{+0.08\,+0.02\,+0.01}$} & \multirow{1}{*}{...} & \multirow{1}{*}{...}\\
 \hline\hline
\end{tabular}\label{tb:dt3}
\end{table}

For those color allowed decay channels, since the
contribution of hard scattering diagrams is highly suppressed by the
Wilson coefficient, the decay amplitude is dominated by the
contribution from factorizable emission diagrams with the Wilson coefficient $a_{1}$, which can be
naively factorized as the product of the Wilson coefficient $a_{1}$,
the decay constant of $D$ meson and the $B$ to tensor meson form
factor. So our predictions basically agree
with the predictions of naive factorization approach in
Ref.\cite{arxiv1010.3077}. The difference is caused by
parameter changes and the interference from suppressed hard scattering and
annihilation type diagrams.

For $B\rightarrow D^{*}(\bar{D}^{*}) T$ decays, we also calculate
the percentage of transverse polarizations, which are listed in Table \ref{tb:dt2} and Table \ref{tb:dt4} (extracted from ref.\cite{zoudt}).
From our computations, we find that some decays disobey the naive factorization assumption.
For those color suppressed $B\rightarrow
\bar{D}^{*}T$ decays with the $\bar{D}^{*}$ emitted, the transverse contributions are dominant  about $70\%$.
 The $\bar{c}$ is right-handed; while the $u$ quark is left-handed, because they are produced through $(V-A)$ current.
 Thus, the $\bar{D}^{*}$ meson is longitudinally polarized.
 But the massive $\bar c$ quark can flip easily from
right handed to left handed. As a result, the polarization of the
$\bar{D}^{*}$ meson becomes transverse with $\lambda=-1$.
On the other hand, because of the additional contribution of orbital angular momentum, the
recoiled tensor meson can also be transversely polarized with $\lambda=-1$ easily.
So the transverse polarization fractions can be as large as 70\%.
While for color suppressed $B\rightarrow {D}^{*}T$ decays with $D^{*}$ meson
emitted, the percentage of transverse polarizations are only at the
range of $20\%$ to $30\%$. the emitted $D^{*}$ meson can also be transversely polarized, but with
 $\lambda=+1$. According to the angular momentum
conservation, the recoiled tensor meson must be also transversely
polarized with $\lambda=+1$. This calls for the tensor
meson getting contributions from both orbital angular momentum and
spin, which is symmetric. Because the    distribution amplitude
 of tensor meson is anti-symmetric, the total wave function is anti-symmetric, which is forbidden  by
 Bose statistics. So the transverse polarization of final states is   suppressed.

\begin{table}
\centering
 \caption{Branching ratios (unit:$10^{-5}$) and the percentage of transverse polarizations $R_{T}$(unit:$\%$) of color-suppressed
 $B_{(s)}\rightarrow \bar{D}^{*}T$ decays in PQCD approach
 together with results from ISGW II model \cite{arxiv1010.3077,prd67014011}.}
\begin{tabular}[t]{lccccc}
\hline\hline
 \vspace{0.2cm}
 \multirow{3}{*}{Decay Modes}& \multirow{3}{*}{Class} & \multicolumn{3}{c}{ \multirow{1}{*}{Branching Ratio($10^{-5}$)}}&
  \multirow{3}{*}{$R_{T}(\%)$}\\
\cline{3-5}
&&\multirow{2}{*}{This Work}&\multirow{2}{*}{SDV[9]}& \multirow{2}{*}{KLO[10]}&\\
&&&&&\\
\hline \vspace{0.13cm}
\multirow{1}{*}{$B^{0}\rightarrow \bar{D}^{*0}a_{2}^{0}$}&\multirow{1}{*}{C} &
\multirow{1}{*}{$39_{-11\,-1\,-1}^{+14\,+2\,+2}$} &\multirow{1}{*}{12}&
\multirow{1}{*}{7.8}& \multirow{1}{*}{$73_{-4\,-8}^{+5\,+9}$}\\
\vspace{0.13cm}
\multirow{1}{*}{$B^{0}\rightarrow \bar{D}^{*0}f_{2}$}& \multirow{1}{*}{C}&
\multirow{1}{*}{$38_{-12\,-1\,-1}^{+14\,+2\,+2}$} & \multirow{1}{*}{13} &
\multirow{1}{*}{8.4}&\multirow{1}{*}{$70_{-6\,-6}^{+6\,+9}$}\\
\vspace{0.13cm}
\multirow{1}{*}{$B^{0}\rightarrow \bar{D}^{*0}K_{2}^{*0}$} & \multirow{1}{*}{C} &
\multirow{1}{*}{$5.3_{-1.4\,-0.7\,-0.2}^{+1.7\,+0.8\,+0.3}$} & \multirow{1}{*}{1.3}&
 \multirow{1}{*}{1.1}&\multirow{1}{*}{$71_{-2\,-9}^{+2\,+9}$}\\
\vspace{0.13cm}
\multirow{1}{*}{$B_{s}\rightarrow \bar{D}^{*0}f_{2}^{\prime}$} & \multirow{1}{*}{C} &
\multirow{1}{*}{$5.0_{-1.7\,-0.5\,-0.2}^{+2.1\,+0.4\,+0.3}$} & \multirow{1}{*}{1.1}&
\multirow{1}{*}{...}&\multirow{1}{*}{$71_{-1\,-7}^{+1\,+8}$}\\
\vspace{0.3cm}
\multirow{1}{*}{$B_{s}\rightarrow \bar{D}^{*0}\bar{K}^{*0}$} & \multirow{1}{*}{C} &
 \multirow{1}{*}{$70_{-24\,-6\,-2}^{+29\,+4\,+4}$} & \multirow{1}{*}{17} & \multirow{1}{*}{...}&
 \multirow{1}{*}{$68_{-1\,-8}^{+2\,+9}$}\\
 \hline\hline
\end{tabular}\label{tb:dt4}
\end{table}

  For the W annihilation(A) type  $B_{(s)}\rightarrow D^{*}T$ decays,
 we also find very large transverse polarizations up to 80\% . The light quark and unti-quark produced
through hard gluon are left-handed or right-handed with equal
opportunity. The $c$ quark is left-handed, and then the $D^{*}$ meson can be longitudinally polarized,
or be transversely polarized with $\lambda=-1$.
For the tensor meson, the anti-quark from weak interaction is
right-handed; while the quark produced from hard gluon can be either
left-handed or right-handed. When taking into account the additional orbital angular momentum, the tensor meson can
be longitudinally polarized or transversely polarized with $\lambda=-1$. So the transverse contributions can become
so large with interference from other diagrams. On the other hand, the W exchange(E) diagrams can not contribute   large transverse contributions, which is consist with the argument in $B \to D^* V$ decays in refs.\cite{prd78014018,zouhaojpg37}.

Some decays involving $f_{2}^{(\prime)}$ in the final states provide a potential way to measure the mixing angle $\theta$ of $f_2$
and $f_2'$. For
example, $B^{0}\rightarrow D^{0}f_{2}^{(\prime)}$,  the branching ratios have
the simple relation
\begin{eqnarray}
r\,=\,\frac{\mathcal {B}(B\rightarrow D f_{2}^{\prime})}{\mathcal
{B}(B\rightarrow D
f_{2})}\,=\,\frac{\sin^{2}\theta}{\cos^{2}\theta},
\end{eqnarray}
since the $s\bar s$ term dose not contribute. In our calculation,  it shows $r \simeq\,0.02  $ with
$\theta=7.8^{\circ}$.

\subsection{$B_c\rightarrow D^{(*)}T$ decays in PQCD approach}

The $B_c$ meson is unique, since it is a heavy quarkonium with two different flavors. Either the heavy $b$ quark or the $c$ quark can decay individually. And the W annihilation diagram decay of the $B_c$ meson is also important due to the large CKM matrix element $V_{cb}$.
Since the $c$ quark in the $B_c\rightarrow D^{(*)}T$ decays is the spectator quark, these decays are dominated by the  $B_c \to  D^{(*)}$ transition form factors with a tensor meson emitted from vacuum, which are prohibited in naive factorization. To our knowledge, these decays are never considered in theoretical papers due to this difficulty of factorization. In order to give the predictions to these decay channels, it is necessary to go beyond the naive factorization to calculate the hard scattering and annihilation-type diagrams. What is more, the annihilation amplitudes will be dominant in these $B_c\rightarrow D^{(*)}T$ decays, because they are proportional to the large CKM matrix elements $V_{cb}  V_{cs(d)}^*$ instead of $V_{ub}  V_{us(d)}^*$  in the emission diagrams.
 Some of these decays are pure W annihilation
type.

 Compared with the $B_c\to D^0 \rho^+$ decay in table III in ref.\cite{pqcd7}, which is dominated by the color allowed
 emission diagram, the corresponding  $B_c\to D^0 a^+_2$ decay has a smaller branching ratio, for the emission of a tensor meson is prohibited.
For  the penguin dominant decay channels $B_c\to D^{0} K^{*+}$ and $D^{+} K^{*0}$, with $b\to s$ transition,
 due to the enhancement of the large CKM elements $V_{cs(d)}$ and large Wilson coefficient $a_1$,
the annihilation diagrams are at the same order magnitude as penguin emission diagrams but with a minus sign shown in table III  of ref.\cite{pqcd7}.
 While  the corresponding $B_c\to D^{0} K_2^{*+}$
 and $D^{+} K_2^{*0}$ decays have much larger branching ratios shown in table~\ref{tb:9},  without the cancellation from emission diagram.  With additional transverse contributions,   the $B_c$ meson decays  to tensor  and  $D^*$ meson final states, have branching ratio as large as $10^{-4}$ in Table~\ref{tb:10} that will be
 easier to search for experiments.

\begin{table}[!h]
\centering
 \caption{Branching ratios ($10^{-6}$) and direct CP asymmetries ($\%$) for some $B_{c}\rightarrow DT$ decays calculated in the PQCD approach .}
\begin{tabular}[t]{l!{\;\;\;\;}c!{\;\;\;\;}c!{\;\;\;\;}c}
\hline\hline

 \multirow{2}{*}{Decay Modes}& \multirow{2}{*}{Class} & \multirow{2}{*}{Br($10^{-6}$)} &
 \multirow{2}{*}{$A_{CP}^{dir}$($\%$)}\\
 &&&\\
 \hline
\vspace{0.13cm}
\multirow{1}{*}{$B_{c}\rightarrow D^{0}a_{2}^{+}$}&\multirow{1}{*}{A} &\multirow{1}{*}{$2.2_{-0.7\,-0.2\,-0.2}^{+0.8\,+0.2\,+0.2}$} &\multirow{1}{*}{$6.47_{-1.15\,-1.59\,-0.74}^{+1.35\,+5.33\,+0.00}$}\\
\vspace{0.13cm}
\multirow{1}{*}{$B_{c}\rightarrow D^{0}K_{2}^{*+}$}& \multirow{1}{*}{A}& \multirow{1}{*}{$31_{-9\,-3\,-1}^{+10\,+3\,+1}$} & \multirow{1}{*}{$-0.44_{-0.15\,-0.22\,-0.02}^{+0.13\,+0.10\,+0.10}$}\\
\vspace{0.13cm}
\multirow{1}{*}{$B_{c}\rightarrow D^{+}K_{2}^{*0}$} & \multirow{1}{*}{A} & \multirow{1}{*}{$32_{-10\,-2\,-1}^{+11\,+3\,+1}$} & \multirow{1}{*}{$0.0$}\\
\vspace{0.3cm}
\multirow{1}{*}{$B_{c}\rightarrow D_{s}^{+}f_{2}^{\prime}$} & \multirow{1}{*}{A} & \multirow{1}{*}{$41_{-11\,-4\,-1}^{+12\,+4\,+1}$} & \multirow{1}{*}{$-0.11_{-0.02\,-0.06\,-0.00}^{+0.02\,+0.03\,+0.02}$}\\
 \hline\hline
\end{tabular}\label{tb:9}
\end{table}
\begin{table}
\centering
 \caption{Branching ratios ($10^{-6}$), direct CP asymmetries ($\%$) and the percentage of transverse polarizations $R_{T}$($\%$) for some  $B_{c}\rightarrow D^{*}T$ decays calculated in the PQCD approach.}
 \vspace{0.2cm}
\begin{tabular}[t]{l!{\;\;}c!{\;\;}c!{\;\;}c!{\;\;}c!{\;\;}c}
\hline\hline

 \multirow{2}{*}{Decay Modes}& \multirow{2}{*}{Class} & \multirow{2}{*}{ Br($10^{-6}$)}& \multirow{2}{*}{$A_{CP}^{dir}$($\%$)}& \multirow{2}{*}{$R_{T}$($\%$)}\\
&&&&\\
 \hline
\vspace{0.18cm}
\multirow{1}{*}{$B_c\rightarrow D^{*0}K_{2}^{*+}$}& \multirow{1}{*}{A}& \multirow{1}{*}{$151_{-27\,-11\,-3}^{+30\,+18\,+5}$} & \multirow{1}{*}{$-0.15_{-0.02\,-0.08\,-0.06}^{+0.02\,+0.05\,+0.03}$} &\multirow{1}{*}{$82.5$}\\
\vspace{0.18cm}
\multirow{1}{*}{$B_c\rightarrow D^{*+}K_{2}^{*0}$} & \multirow{1}{*}{A} & \multirow{1}{*}{$158_{-29\,-15\,-13}^{+31\,+16\,+0}$} & \multirow{1}{*}{$0.0$}& \multirow{1}{*}{$80.3$}\\
\vspace{0.18cm}
\multirow{1}{*}{$B_c\rightarrow D_{s}^{*+}\bar{K}_{2}^{*0}$} & \multirow{1}{*}{A} & \multirow{1}{*}{$8.9_{-1.6\,-0.9\,-0.3}^{+1.7\,+0.8\,+0.5}$} & \multirow{1}{*}{$2.3_{-0.1\,-0.5\,-0.0}^{+0.2\,+0.9\,+0.0}$} & \multirow{1}{*}{$82.0$}\\
\vspace{0.36cm}
\multirow{1}{*}{$B_c\rightarrow D_{s}^{*+}f_{2}^{\prime}$} & \multirow{1}{*}{A} & \multirow{1}{*}{$190_{-28\,-13\,-4}^{+31\,+20\,+6}$} & \multirow{1}{*}{$-0.036_{-0.003\,-0.012\,-0.001}^{+0.004\,+0.011\,+0.008}$} & \multirow{1}{*}{$89.5$}\\
 \hline\hline
\end{tabular}\label{tb:10}
\end{table}

The LHC experiment, specifically the LHCb, can produce around
$5 \times 10^{10}$ $B_c$ events each year\cite{haochu1,haochu2}.
The $B_c$ decays with a decay rate  at the level of $10^{-6}$ can be
detected with a good precision at LHC experiments \cite{su}. So some of these
$B_c\rightarrow D^{(*)}T$ decays can be observed in the experiments. For example,
$B_{c} \to D^{(*)+}K_{2}^{*0}$, the branching ratio is at the order of  $ 10^{-5}$($ 10^{-4}$). Taking into account
 the branching ratios of $D^{+}$ and $K_{2}^{*0}$ decays with charged final states
 ($10\%$ ($D^{+}\to K^{-}\pi^{+}\pi^{+}$)\cite{gaoyuanning}
and $25\%$ ($\mathcal{B}(K_{2}^{*0}\to K\pi)=(49.9\pm1.2)\%$) respectively)
and assuming a total efficiency of $1\%$ \cite{gaoyuanning}, one can expect about dozens of events every year.
For $B_c\to D_{s} f_{2}^{\prime}$ with branching ratio $4\times 10^{-5}$, taking into account
the branching ratios of $D_{s}^{+}$ and $f_{2}^{\prime}$ decays with charged final states (6\% ($D_{s}^{+}\to K^{+}K^{-}\pi^{+}$)
and $45\%$ ($\mathcal{B}(f_{2}^{\prime}\to K\bar{K})
=(88.7\pm2.2)\%$) respectively) and assuming a total efficiency of $1\%$, one can expect about one hundred
events every year. They are the most promising channels to be measured in the current experiments.

Due to the same reason in  W annihilation type $B_{(s)}\rightarrow D^{*}T$ decays, for the W annihilation diagrams dominant decays,
we also predict the percentages of the transverse polarization around $80\%$ or even bigger.
The CP averaged branching ratios, direct CP asymmetries, and the transverse polarization fractions for all the  decays of the type $B_{(s)}\rightarrow D^{*}T$ decays  can be found in
 ref \cite{zoubcdt}.

\section{SUMMARY}

We have studied the $B_{u,d,s,c}$ decays involving a light tensor meson in final states within the framework of perturbative
QCD approach. We  calculate the contributions of
different diagrams, especially the hard scattering and annihilation type diagrams, which are important
to explain the large experimental data and the large transverse polarization fractions.
For some decays with tensor meson emitted or pure annihilation type decays, we give the predictions for
the first time. For those penguin dominant decays and color-suppressed decays, we give larger and
more reliable predictions, which agree the experimental data better. For those color suppressed
$B_{(s)}\rightarrow \bar{D}^{*}T$ decays, the transversely
polarized contributions from hard scattering diagrams are very
large. For those W annihilation type $B_{(c)}\rightarrow D^{*}T$ decays,
the transverse polarized contributions from factorizable
annihilation diagrams are  as large as $80\%$.

\section*{Acknowledgment}

We are grateful to Xin Yu, Qin Qin and H.-n. Li for collaborations.
The work of   is supported in part by the Foundation of Yantai University under
Grant No. WL07052 and the National Science Foundation of China under Grant Nos. 11375208,
11228512 and 11235005.

\end{document}